\documentclass[aps,amsmath,amssymb,twocolumn,showpacs,superscriptaddress,prb]{revtex4} 
\usepackage{graphicx}
\newcommand{\bea}{\begin{eqnarray}} 
\newcommand{\eea}{\end{eqnarray}} 
\newcommand{\be}{\begin{equation}} 
\newcommand{\ee}{\end{equation}} 
 
\newcommand{\beq}{\begin{eqnarray}} 
\newcommand{\eeq}{\end{eqnarray}}

\def\H1{\widehat{H}_1}

\newcommand{\up}{\uparrow} 
\newcommand{\down}{\downarrow} 
\newcommand{\la}{\langle} 
\newcommand{\ra}{\rangle}

\newcommand{\bS}{{\bf S}}

\begin{document}

\title{Magnetization plateau in the S=$\frac{1}{2}$ spin ladder with alternating rung exchange}

\author{G.I. Japaridze\footnote{Permanent address: Andronikashvili Institute  of  Physics, Tamarashvili 6, 0177 Tbilisi,  Georgia.} }

\affiliation{D\'{e}partement de Physique, Universit\'{e} de Fribourg, 
CH-1700, Fribourg, Switzerland}

\author{E. Pogosyan}

\affiliation{Andronikashvili Institute of Physics, Tamarashvili 6, 0177 Tbilisi, Georgia}

\begin{abstract} 

We have studied the ground state phase diagram of a spin ladder with alternating rung exchange $J^{n}_{\perp} = J_{\perp}\left[1 + (-1)^{n} \delta \right]$ in a magnetic filed, in the limit where the rung coupling is dominant. In this limit the model is mapped onto an $XXZ$ Heisenberg chain in a uniform and staggered longitudinal magnetic fields, where the amplitude of the staggered field is $\sim \delta$. We have shown that the magnetization curve of the system exhibits a plateau at magnetization equal to the half of the saturation value. The width of a plateau scales as $\delta^{\nu}$, where $\nu =4/5$ in the case of ladder with isotropic antiferromagnetic legs and $\nu =2$ in the case of ladder with isotropic ferromagnetic legs. We have calculated four critical fields ($H^{\pm}_{c1}$ and $H^{\pm}_{c2}$) corresponding to transitions between different magnetic phases of the system. We have shown that these transitions belong to the universality class of the commensurate-incommensurate transition.

\end{abstract}
\maketitle

\section{Introduction}

A theoretical understanding of the magnetic properties of quantum spin systems, in particular of spin $S=1/2$ {\em isotropic antiferromagnetic} two-leg ladders, has attracted a lot of interest for a number of reasons. On the one hand, there was remarkable progress in recent years in the fabrication of such ladder compounds. \cite{RiceDagotto} On the other hand, spin-ladder models pose interesting theoretical problems since antiferromagnetic two-leg ladder systems have a gap in the excitation spectrum and, in the presence of a magnetic field, they reveal an extremely rich behavior, dominated by quantum effects. These quantum phase transitions were intensively investigated both theoretically$^{2-14}$ and experimentally.$^{15-20}$

The discovery of a magnetic field induced gap in the $Cu$-bensoate \cite{Cu-Benzoate_Exp_1} and other spin chain materials\cite{Cu-Benzoate_Exp_2} 
have increased the interest for magnetic quantum phase transitions which are determined by the combined effects of the uniform and staggered components of the effective magnetic field. \cite{CuBen_Affleck_Oshikawa,CuBen_Essler_Tsvelik,Huang_Affleck,Mila_04,Kolezhuk_2004,DK_04,Wang_1b,Wang_1c}

\begin{figure}[tb]
\vspace{5mm}
\includegraphics[width=75mm]{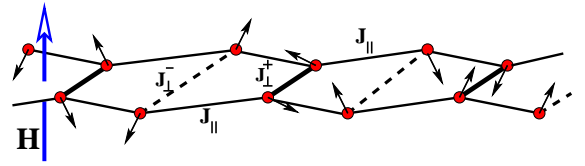}
\vspace{5mm}
\caption{The ladder with alternating rung exchange }
\label{Fig:Fig1}
\end{figure}

In this paper we study the ground state magnetic phase diagram of the spin $S=1/2$ two-leg ladder with alternating rung exchange given by the Hamiltonian (Fig. \ref{Fig:Fig1})
\begin{eqnarray} 
{\cal H} &=& J_{\parallel} \sum_{n,\alpha} \bS_{n,\alpha} \cdot \bS_{n+1,\alpha}
- H  \sum_{n,\alpha} S^{z}_{n,\alpha} \nonumber\\
& + & J_{\perp} \sum_{n} \left[1 + (-1)^{n} \delta \right] 
\bS_{n,1} \cdot \bS_{n,2}\, ,
\label{FL_4_Hamiltonian} 
\end{eqnarray} 
where $\bS_{n,\alpha}$ is a spin $S=1/2$ operator of rung n (n=1,...,N) and leg $\alpha$ ($\alpha=1,2$). The interleg coupling is antiferromagnetic, $J^{\pm}_{\perp} = J_{\perp}(1 \pm \delta) > 0 $.

\begin{figure}[tb]
\vspace{5mm}
\includegraphics[width=55mm]{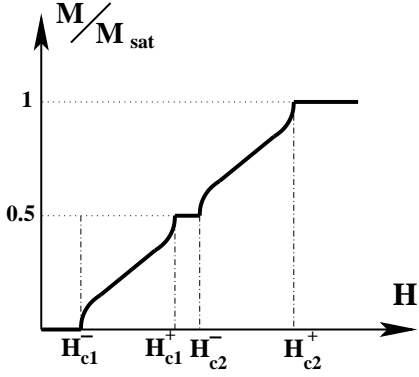}
\vspace{0mm}
\caption{Schematic drawing of the magnetization (in units of saturated magnetization $M_{sat}$) of a two-leg isotropic ladder with alternating rungs as a function of the external magnetic field.}
\label{Fig:Fig2}
\end{figure}

We restrict our consideration to the limit of strong rung exchange $J^{\pm}_{\perp} \gg |J_{\parallel}|,\delta J_{\perp}$ and map the model onto spin-1/2 $XXZ$ Heisenberg chain in the presence of both longitudinal uniform and staggered magnetic fields, with the amplitude of the staggered component of the magnetic field proportional to $\sim \delta J_{\perp}$. We study the ground state phase diagram of the effective spin-chain model and show, that the alternation of the rung-exchange leads to the dynamical generation of a new energy scale in the system and to the appearance of two additional quantum phase transitions in the magnetic ground state phase diagram. These transitions manifest themselves most clearly in the presence of a new magnetization plateau at magnetization equal to one half of its saturation value (see Fig.\ref{Fig:Fig2}). The magnetic phase diagram is characterized by the following four critical fields: the field $H^{-}_{c1}$, which corresponds the the transition from a gapped rung-singlet phase to the gapless paramagnetic phase; the critical fields $H^{+}_{c1}$ and $H^{-}_{c2}$ which mark end-points of the magnetization plateau and the saturation field $H^{+}_{c2}$.  The width of the plateau scales as $\delta^{\nu}$, where $\nu =4/5$ in the case of a ladder with isotropic antiferromagnetic legs and $\nu =2$ in the case of a ladder with isotropic ferromagnetic legs. Therefore this magnetic phase diagram is generic for a standard isotropic ladder with alternating rung exchange. However, in the case of a ladder with ferromagnetic legs and frustrating diagonal interleg exchange, the intermediate magnetization plateau dissappears for sufficiently strong ferromagnetic diagonal coupling.  

\section{Derivation of the effective Hamiltonian}

In this section we derive the effective spin-chain model to describe the strong rung-exchange limit $J_{\perp} \gg (\delta J_{\perp}), |J_{\parallel}|$ of the model (\ref{FL_4_Hamiltonian}). To obtain the spin chain Hamiltonian we follow the route already used to study the standard ladder models in the same limit of strong rung exchange.\cite{Mila_98,Totsuka_98}
We start from the case $J_{\parallel} = 0$. In this limit the system decouples into a  set of noninteracting rungs with couplings $J^{+}_{\perp}$ and $J^{-}_{\perp}$. In this case, an eigenstate of ${\cal H}$ is written as a product of rung states. At each rung two spins $\bS_{n,l}$ and $\bS_{n,2}$ are either in a singlet state $|s^{0}_{n}\ra = \frac{1}{\sqrt{2}}(|\up\down\ra - |\down\up\ra)$ or in one of the triplet states: $|t^{+}_{n}\ra  = |\up\up\ra\, ,|t^{-}_{n}\ra  = |\down \down\ra$ and $|t^{0}_{n}\ra = \frac{1}{\sqrt{2}}(|\up\down\ra + |\down\up\ra)$. Their energies are respectively: $E(s^{0}_{n}) = -3J^{n}_{\perp}/4$, $E(t^{+}_{n}) = J^{n}_{\perp}/4 - H$, $E(t^{0}_{n}) = J^{n}_{\perp}/4$ and $E(t^{-}_{n})=J^{n}_{\perp}/4 + H$, where 
$J^{n}_{\perp}=J_{\perp}[1 + (-1)^{n}\delta]$ 

When $H$ is small, the ground state consists of a product of rung singlet. As the field $H$ increases, the energy of the state $|t^{+}_{2n-1} \rangle$ decreases and at $H = H_{c1} = J^{-}_{\perp}$ this state is degenerate with $|s^{0}_{2n-1}\rangle$. 
Thus, at $H = H_{c1}$ the ground state of a given odd rung undergoes a transition from the singlet $|s^{0}_{2n-1}\ra$ to the triplet $|t^{+}_{2n-1}\ra $ and the total magnetization of the system jumps discontinuously from zero to $0.5M_{sat}=N/2$. With further increase of the magnetic field, at $H_{c1} < H < H_{c2}$ the magnetization remains constant. However since for $H_{c2} > J^{+}_{\perp}$ the energy of the state $|t^{+}_{2n} \rangle $ is lower then $|s^{0}_{2n}\rangle$, the magnetization once again increases discontinuously from $0.5M_{sat}$ to $M_{sat}=N$ for $H_{c2} = J^{+}_{\perp}$.

For $J_{\parallel} \neq 0$ these abrupt transitions are broadened into intervals $H^{-}_{c1}<H<H^{+}_{c1}$ and $H^{-}_{c2}<H<H^{+}_{c2}$, respectively. Two different scenarios are possible. Either $H^{+}_{c1} < H^{-}_{c2}$ and the magnetization plateau with $M = 0.5M_{sat}$ remains. Or $H^{+}_{c1} > H^{-}_{c2}$ and  alternation of the rung exchange is irrelevant. In the latter case, the model shows a similar behavior as the standard two-leg ladder in a magnetic field: with increasing magnetic field in the range between $H^{-}_{c1}$ and $H^{+}_{c2}$ the magnetization continuously evolves from a nonmagnetic phase at $H \leq H_{c1}^{-}$ into the fully polarized ferromagnetic state at $H \geq H_{c2}^{+}$. 

The easiest way to obtain the effective model is to split the Hamiltonian (\ref{FL_4_Hamiltonian}) into three parts:
$$
H = H_{0}^{(o)} + H_{0}^{(e)} + H_{int} 
$$
\begin{eqnarray} 
H_{0}^{(o)} &=&  J_{\perp}^{(-)}\sum_{n}\bS_{2n-1,1}\cdot \bS_{2n-1,2}\nonumber\\ 
&-& H_{c1}\sum_{n,\alpha}\,S_{2n-1,\alpha}^{z} \, ,\\
\label{H_0_odd} 
H_{0}^{(e)} &=&  J_{\perp}^{(+)}\sum_{n=1}^{N/2}\bS_{2n,1}\cdot \bS_{2n,2} \nonumber\\ 
&-& H_{c2}\sum_{n,\alpha}\,S_{2n,\alpha}^{z} 
\label{H_0_even}
\end{eqnarray}
and
\bea
H_{int}  &=& J_{\parallel} \sum_{n} \sum_{\alpha} \bS_{n,\alpha}\bS_{n+1,\alpha}\nonumber\\ 
&-&\left(H-H_{c1}\right)\sum_{n,\alpha}S_{2n-1,\alpha}^{z}\nonumber \\ 
&-&\left(H-H_{c2}\right)\sum_{n,\alpha}S_{2n,\alpha}^{z} \, .
\label{Int} 
\eea
The ground state of ${\cal H}_0$ is $2^N$ times degenerate, since each rung 
can be in the state $|\,s_{0}\ra$ or $|t_{+}\ra$ and the first excited state has an energy of the order of $J_\perp$. ${\cal H}_{int}$ will lift the degeneracy in the ground state manifold, leading to an effective Hamiltonian that can be derived by standard perturbation theory.\cite{Mila_98} 

Let us start by introducing pseudo-spin $\tau=1/2$ operators, ${\mbox{\boldmath $\tau$}}_{n}$ which act on these states as
\begin{eqnarray} 
&&\tau_{n}^{z}|\,s_{0}>_{n}~ = -\frac{1}{2}|\,s_{0}>_{n}\, , ~~~~
\tau_{n}^{z}|\,t_{+}>_{n} ~ = \frac{1}{2}|t_{+}>_{n}\, ,\nonumber \\ 
&&\tau_{n}^{+}|\,s_{0}>_{n} ~ = ~~~|\,t_{+}>_{n}\, , ~~~~~ 
\tau_{n}^{+}|t_{+}>_{n}~ = ~~~0 \, , \\ 
&&\tau_{n}^{-}|\,s_{0}>_{n} ~ = ~~~~ 0 \, ,~~~~~~~~~~~~ 
\tau_{n}^{-}|\,t_{+}>_{n} = |\,s_{0}>_{n}\, . \nonumber 
\end{eqnarray} 
The relation between the real spin operator $\bS_{n}$ and the pseudo-spin operator ${\mbox{\boldmath $\tau$}}_{n}$ in this restricted subspace can be easily derived by inspection,
\be
S^{\pm}_{n,\alpha} = (-1)^{\alpha}\frac{1}{\sqrt{2}}\tau^{\pm}_{n}\, , \quad  
S^{z}_{n,\alpha} = \frac{1}{2}\left(\frac{1}{2}+\tau^{z}_{n}\right) \, .
\label{relations} 
\ee
Using (\ref{relations}), to first order and up to constant, we easily obtain the effective Hamiltonian 
\begin{eqnarray} 
H_{eff}& = & \sum_{n} \{J_{xy}\left(\tau_{n}^{x} \tau_{n+1}^{x} + 
\tau_{n}^{y} \tau_{n+1}^{y}\right) + J_{z} \tau_{n}^{z} \tau_{n+1}^{z}\} \nonumber \\ 
& - & h_{eff}^{0}\sum_{n}\tau_{n}^{z}  - h_{eff}^{1}\sum_{n}(-1)^{n}\tau_{n}^{z} \, , 
\label{FL_4_Hamiltonian_XYZ}
\end{eqnarray} 
where 
\begin{eqnarray} 
J_{xy} & = & J_{\parallel}\, ,\quad J_{z} = \frac{1}{2}J_{\parallel}\, , \label{JxyJz}\\
h_{eff}^{0} & = & H - J_{\perp} - \frac{J_{\parallel}}{2}\, ,  \\ 
h_{eff}^{1} & = & \delta J_{\perp}\, .
\label{eff} 
\end{eqnarray} 
Thus the effective Hamiltonian is nothing but the $XXZ$ Heisenberg chain, with anisotropy $J_{z}/J_{xy} \equiv \Delta =1/2$ in uniform and staggered longitudinal magnetic fields. It is worth to notice that a similar problem has been studied intensively in recent years.
\cite{CuBen_Affleck_Oshikawa,CuBen_Essler_Tsvelik,Huang_Affleck,Mila_04,Kolezhuk_2004,DK_04,Wang_1b,Wang_1c}

\section{Magnetic phase diagram}

\subsection{The first critical field $H^{-}_{c1}$ and the saturation field $H^{+}_{c2}$}

We first calculate of the critical field $H^{-}_{c1}$, corresponding to the transition from a gapped rung-singlet phase to a gapless paramagnetic phase, and the saturation field $H^{+}_{c2}$. 

For $H < H^{-}_{c1}$ the ground state of the system corresponds to the gapped rung singlet phase with zero magnetization. For $H > H^{+}_{c2}$ the system is in the fully polarized ferromagnetic phase. The easiest way to express $H^{-}_{c1}$ and $H^{+}_{c2}$ in terms of ladder parameters $J_{\parallel}$, $J_\perp$ and $\delta$ is to perform the Jordan-Wigner transformation which maps the problem onto a system of interacting spinless fermions:
\begin{eqnarray}
H_{sf} & = & t\sum_{n}(a_{n}^{+}a_{n+1} + h.c.) + V \sum_{n}\rho_{n}\rho_{n+1} \nonumber \\
&-& \sum_{n}\big[\mu_{0} + (-1)^{n}\mu_{1} \big]\rho_{n}
\label{Hamiltonian_SpFrm}
\end{eqnarray}
where
\begin{eqnarray} 
t & = & \frac{1}{2}J_{\parallel}\, , \qquad ~~~~~~~ V = \frac{1}{2}J_{\parallel} \\ 
\mu_{0}& = & \frac{1}{2}J_{\parallel} + h_{eff}^{0} \, , \quad 
\mu_{1}= h_{eff}^{1}\, .
\end{eqnarray} 
The lowest critical field $H^{-}_{c1}$ corresponds to that value of the chemical potential $\mu_{0c}$ for which the band of spinless fermions starts to fill up. In this limit we can neglect the interaction term in Eq. (\ref{Hamiltonian_SpFrm}) and obtain the model of free massive particles with spectrum 
\begin{equation}
E^{\pm}(k)=-\mu_{0}  \pm \sqrt{ J_{\parallel}^{2} cos^{2}(k) + \mu_{1}^{2}}\, .
\label{energy}
\end{equation}
The chemical potential corresponding to $H^{-}_{c1}$ is given by $\mu_{0c}=-\sqrt{J^{2}+\mu_{1}^{2}}$, i.e.
\begin{equation}
H_{c1}^{-} = J_{\perp} - \sqrt{J_{\parallel}^{2} + (\delta J_{\perp})^{2}}\, .
\label{H^{-}_{c1}}
\end{equation}

A similar argument can be used to determine $H^{+}_{c2}$. It is useful to make a  particle-hole transformation and estimate $H^{+}_{c2}$ from the condition where the transformed hole band starts to fill. This gives
\begin{equation}
H^{+}_{c2} = J_{\perp} + J_{\parallel} + \sqrt{J_{\parallel}^{2} + (\delta J_{\perp})^{2}}\, .
\end{equation}

\subsection{Magnetization plateau: $H^{+}_{c1}$ and $H^{-}_{c2}$ }

To determine the values of the remaining two critical fields $H^{+}_{c1}$ and $H^{-}_{c2}$ we consider the model (\ref{FL_4_Hamiltonian_XYZ})for $h_{eff}^{0},\, h_{eff}^{1} \ll J_{\parallel}$. 

For $h_{eff}^{0}=h_{eff}^{1}=0$ the Hamiltonian (\ref{FL_4_Hamiltonian_XYZ}) with  anisotropy parameter $|\Delta| < 1$ is known to be critical. The long wavelength excitations are described by the standard Gaussian theory with Hamiltonian\cite{LP}
\begin{equation}\label{SpinChainBosHam}
{\cal H}_{leg} =  \int dx \, \frac{v_{s}}{2}[(\partial_x \phi)^{2} +  (\partial_x \theta)^{2}]\, .
\end{equation}
Here $\phi(x)$ and $\theta(x)$ are dual bosonic fields, $\partial_t \phi =
v_{s} \partial_x \theta $, and satisfy the following commutational relation
\begin{eqnarray}
\label{regcom}
&& [\phi(x),\theta(y)]  = i\Theta (y-x)\,,  \nonumber\\ 
&& [\phi(x),\theta(x)]  =i/2\, .
\end{eqnarray}
The velocity of spin excitation $v_{s}$ is fixed 
from the Bethe ansatz solution as
\bea
v_{s} &=& J_{\parallel}\frac{K}{2K-1}\sin{\left(\pi/2K\right)}\,,
\label{u}
\eea
where the spin-stiffness parameter $K$ is given by
\begin{equation}
K=\frac{1}{2\left(1-\frac{1}{\pi}\arccos\Delta \right)} \, .
\label{K}
\end{equation}
Thus the parameter $K$ increases monotonically along the $XXZ$
critical line $ -1 < \Delta < 1 $ from its minimal value $K=1/2$ at
$\Delta =1$ (isotropic antiferromagnetic chain), to unity at
$\Delta =0$ (the $XY$ chain) and diverges at the ferromagnetic instability 
point of a single chain $\Delta = -1$. 

To obtain the continuum version of the Hamiltonian (\ref{FL_4_Hamiltonian_XYZ}) we use the standard bosonization expression of the spin operator\cite{GNT}
\be
\tau_{n}^{z}  =   \sqrt{\frac{K}{\pi}} \partial_x \phi \, + \nonumber\\ 
(-1)^n \, \frac{A}{\pi} \sin(\sqrt{4\pi K}\phi) \, ,
\label{bosforSz}
\ee
where $A$ is a non-universal real constant of the order of unity\cite{Hikihara}, and get the continuum Hamiltonian
\begin{eqnarray}
H_{Bos} &=& \int dx \Big\{ \frac{v_{s}}{2}[(\partial_{x}\phi)^{2} + (\partial_x\theta)^{2} ] \nonumber\\
 & + & \frac{h_{eff}^{1}}{\pi a_{0}} \sin(\sqrt{4\pi K}\phi) - h_{eff}^{0} \sqrt{\frac{K}{\pi}}\partial_{x}\phi\Big\}\, .
\label{FL_4_Bos_Hamiltonian}
\end{eqnarray}

The Hamiltonian (\ref{FL_4_Bos_Hamiltonian}) is the standard Hamiltonian for 
the commensurate-incommensurate transition which has been intensively studied in the past using bosonization\cite{C_IC_transition} and the Bethe ansatz. \cite{JNW_1984}  Below we use these results to describe the magnetization plateau and the transitons from a gapped (plateau) to gapless paramagnetic phases. 

Let us first consider $h_{eff}^{0}=0$. In this case the continuum theory of the initial ladder model in the magnetic field $H=J_{\perp} +J_{\parallel}/2$ is given by the quantum sine-Gordon (SG) model with a massive term $\sim h_{eff}^{1}\sin(\sqrt{4\pi K}\phi)$. From the exact solution of the SG model\cite{DHN} it is known that the excitation spectrum is gapless for $K \geq 2$ and has a gap in the interval $0 < K < 2$. At $K=1$ the sine-Gordon model is equivalent to the theory of free massive fermions with $m=h_{eff}^{1}$. At $1 < K < 2$ the excitation spectrum of the model consists of solitons and antisolitons with mass $M$, while for $0 < K < 1$ the spectrum contains also  soliton-antisoliton bound states ("breathers"). The exact relation between the soliton mass $M$ and the bare mass  $h_{eff}^{1}=\delta J_{\perp}$ is given by\cite{Zamolodchikov_95}
\be
M = J_{\parallel}{\cal C}(K) \left(\delta J_{\perp}/J_{\parallel}\right)^{1/(2-K)} \, ,
\label{SG-mass_Zamolodchikov}
\ee
where
\be
{\cal C}(K) = \frac{2\Gamma(\frac{1}{2\nu})}{\sqrt{\pi}\Gamma(\frac{1}{2} +\frac{1}{2\nu})}\cdot
\left[\frac{\Gamma(1-K/2)}{2\Gamma(K/2)} \right]^{1/(2-K)}\, . 
\ee

It is straightforward to get from (\ref{K}), that at $\Delta=1/2$ the spin stiffness parameter $K=3/4$. Therefore, for $h_{eff}^{0}=0$  the sine-Gordon Hamiltonian (\ref{FL_4_Bos_Hamiltonian}) is in the strong coupling (massive) regime. In this case the low-energy behavior of the system is determined by the strongly relevant staggered magnetic field (i.e. alternating part of the rung exchange), represented by the term $h_{eff}^{1}\sin(\sqrt{3\pi}\phi)$. In the ground state the field $\phi$ is pinned in one of the minima of the staggered field potential 
\be
\la 0| \sqrt{3\pi}\phi|0 \ra = -\pi/2 + 2\pi n \, .
\label{minima}
\ee
In view of (\ref{bosforSz}) we conclude that this state corresponds to a  long-range-ordered  antiferromagnetic phase of the effective Heisenberg chain (\ref{FL_4_Hamiltonian_XYZ}), i.e. to a phase of the initial ladder system, where  odd rungs have a dominant triplet character and even rungs are predominantly singlets. 

At $h_{eff}^{0} \neq 0$ (i.e. $H \neq J_{\perp} + J_{\parallel}/2$) the very presence of the gradient term in the Hamiltonian (\ref{FL_4_Bos_Hamiltonian}) makes it necessary to consider the ground state of the sine-Gordon model in sectors with nonzero topological charge. The effective chemical potential $\sim h_{eff}^{0}\sqrt{\frac{K}{\pi}}\partial_{x}\phi$ tends to change the number of particles in the ground state i.e. to create a finite and uniform density solitons. It is clear that the gradient term in (\ref{FL_4_Bos_Hamiltonian}) can be eliminated by a gauge transformation $\phi \rightarrow \phi_{s} + h_{eff}^{0}\sqrt{\frac{K}{\pi}}\,x$, however this immediately implies that the vacuum distribution of the fieled $\phi$ will be shifted with respect of the minima (\ref{minima}). This  competition between contributions of the smooth and staggered components of magnetic field is resolved as a continuous phase transition from a gapped state at $|h_{eff}^{0}| < M$ to a gapless (paramagnetic) phase at $|h_{eff}^{0}| > M$, where $M$ is the soliton mass.\cite{C_IC_transition} 

For our effective Hamiltonian (\ref{FL_4_Hamiltonian_XYZ}) with $\Delta=1/2$, the spin stiffness parameter $K$ is 3/4 model (Eq.(\ref{K})) and the commensurate-incommensurate transition in the effective sine-Gordon theory at $\pm h_{eff}^{0} = M$ gives two additional critical values of the magnetic field
\begin{equation}
H_{c1}^{+} = J_{\perp} + J_{\parallel}/2 - J_{\parallel}
{\cal C}_{0}\left(\delta J_{\perp}/J_{\parallel} \right)^{4/5}
\label{Hc1+}
\end{equation}
and 
\begin{equation}
H_{c2}^{-} = J_{\perp} + J_{\parallel}/2 + J_{\parallel}
{\cal C}_{0}\left(\delta J_{\perp}/J_{\parallel}\right)^{4/5}\, ,
\label{Hc1+}
\end{equation}
where ${\cal C}_{0}={\cal C}(3/4)=1.11428$.

As usual in the case of quantum commensurate-incommensurate transition transitions, the magnetic susceptibility of the system shows a square-root divergence at the transition points: 
$$
\label{Susceptibility} 
\chi(H) = \left\{ \begin{array}{l@{\quad}}\left(H_{c1}^{+} - H \right)^{-1/2}\, \hspace{0.3cm} {\mbox for}\,~~~\hspace{0.5cm} H < H_{c1}^{+}
\\
\vspace{0.7mm}
~~~~~~~ 0 \hskip0.6cm \, ~~~~~~~~~ {\mbox for}\,\hskip0.5cm H_{c1}^{+} < H < H_{c2}^{-}\\
\left(H-H_{c2}^{-} \right)^{-1/2}\, \hskip0.3cm {\mbox for}\,~~~\hskip0.5cm  H > H_{c2}^{-}
\end{array}\right. \, . 
$$

\subsection{Magnetic phase diagram}

Summarizing the results of the previous subsections we obtain the following magnetic phase diagram for a ladder with alternating rung exchange (see Fig. \ref{Fig:Fig2}).  For $H<H_{c1}^{-}$, the system is in a rung-singlet phase with zero magnetization and vanishing magnetic susceptibility. For $H > H_{c1}^{-}$ some of singlet rungs melt and the magnetization increase as $\left(H-H_{c1}^{-} \right)^{1/2}$. With further increase of the magnetic field the system gradually crosses to a regime with linearly increasing magnetization. However, in the vicinity of the magnetization plateau, for  $H \leq H_{c1}^{+}$ this linear dependence changes and the magnetization once again shows a square-root behavior $M - \frac{1}{2}M_{sat} \sim  - \left(H_{c1}^{+} - H \right)^{1/2}$. For fields in the interval between $H_{c1}^{+}$ and $H_{c2}^{-}$ the magnetization is constant $M = 0.5 M_{sat}$.  At $H > H_{c2}^{-}$ the magnetization increases as $M \sim 0.5 M_{sat} + \left(H-H_{c2}^{-}\right)^{1/2}$, then passes again through a linear regime until, in the vicinity of the saturation field $H_{c2}^{+}$, it becomes $M \sim M_{sat} - \left(H_{c2}^{+} - H \right)^{1/2}$. 

The width of the magnetization plateau at $M = 0.5 M_{sat}$ is given by
\be
H_{c2}^{-}- H_{c1}^{+} \simeq 2J_{\parallel}\left(\delta J_{\perp}/J_{\parallel} \right)^{4/5}
\, .
\label{Platou}
\ee

\subsection{Ladder with ferromagnetic legs}

The existence of a magnetization plateau at $M=0.5M_{sat}$ is not limited to the ladder with antiferromagnetic exchange, but is also found for a ladder with {\em isotropic ferromagnetic} legs ($J_{\parallel}=-|J_{\parallel}|<0$) coupled by an antiferromagnetic rung exchange ($J_{\perp} > 0$). In this case the effective spin chain model is also given by the Hamiltonian (\ref{FL_4_Hamiltonian_XYZ}), but with different parameters:
\begin{eqnarray} 
J_{xy} & = & |J_{\parallel}|\, ,\quad J_{z} = -\frac{1}{2}|J_{\parallel}|\, , \label{JxyJz}\\
h_{eff}^{0} & = & H - J_{\perp} + \frac{1}{2}|J_{\parallel}|\, ,  \\ 
h_{eff}^{1} & = & \delta J_{\perp}\, .
\label{eff_2} 
\end{eqnarray} 
Thus, in the case of ferromagnetic legs the anisotropy parameter of the effective $XXZ$ is $\Delta =-1/2$ and consequently the spin stiffness parameter $K$ is given by $K=3/2$. The equivalent sine-Gordon theory (\ref{FL_4_Bos_Hamiltonian}) with a massive term  $\sim h_{eff}^{1}\sin(\sqrt{6\pi}\phi)$ remains in the gapped strong-coupling regime. Using Eq. (\ref{SG-mass_Zamolodchikov}) we find an excitation gap $M \sim |J_{\parallel}|\left(\delta J_{\perp}/|J_{\parallel}| \right)^{2}$. Correspondingly the width of the magnetization plateau in this case equals
\be
H_{c2}^{-}- H_{c1}^{+} \simeq 2J_{\parallel}\left(\delta J_{\perp}/J_{\parallel} \right)^{2}\, .
\label{Platou_2}
\ee

\section{Generalized ladder}

In this section we consider a generalized ladder model with frustrating (diagonal) interleg interactions. The Hamiltonian is given by
\begin{eqnarray} 
{\cal H} & = &  J_{\parallel} \sum_{n,\alpha} \bS_{n,\alpha} \cdot \bS_{n+1,\alpha}
+ H  \sum_{n,\alpha} S^{z}_{n,\alpha} \nonumber\\
& + &J^{\prime}_{\perp} \sum_{n} \left(\, \bS_{n,1} \cdot \bS_{n+1,2} + \bS_{n,2} \cdot \bS_{n+1,1} \, \right)
\nonumber\\ 
& + & J_{\perp} \sum_{n} \left(1 + (-1)^{n} \delta \right) 
\bS_{n,1} \cdot \bS_{n,2}\, .
\label{FL_4_Hamiltonian_Ext} 
\end{eqnarray} 
Lat us first consider the case of antiferromagnetic legs ($J_{\parallel}>0$). Assuming $J_{\perp} \gg J_{\parallel}, |J^{\prime}_{\perp}|, \delta J_{\perp}$ we easily obtain the parameters of $H_{eff}$, 
\begin{eqnarray} 
\Delta & = & \frac{1}{2}\frac{J_{\parallel} + J^{\prime}_{\perp}}{J_{\parallel} - J^{\prime}_{\perp}}\, ,\\
h_{eff}^{0} & = & H - J_{\perp} - \frac{J_{\parallel}}{2} - \frac{J^{\prime}_{\perp}}{2}\, ,  \\ 
h_{eff}^{1} & = & \delta J_{\perp}\, .
\label{eff} 
\end{eqnarray} 
Below we briefly discuss following limiting cases: (i) $K=1/2$ ($\Delta=1$), (ii) $K=1$ ($\Delta=0$) and (iii) $K=2$ ($\Delta=-1/\sqrt{2}$). 

The case (i) is reached for antiferromagnetic diagonal exchange $J^{\prime}_{\perp} = 1/3J_{\parallel}$. In this case the width of the magnetization plateau is
\be
H_{c2}^{-}- H_{c1}^{+} \simeq 2J_{\parallel}\left(\delta J_{\perp}/J_{\parallel} \right)^{2/3}\, .
\label{Platou_2}
\ee

The case (ii) is reached for ferromagnetic diagonal exchange $J^{\prime}_{\perp} =-J_{\parallel}$, where the effective continuum theory becomes the theory of free massive fermions. At $K=1$ the spin gap at $h_{eff}^{0}=0$ is 
\be
M = \delta J_{\perp} \, .
\label{SG-mass_Zamolodchikov_2}
\ee
Correspondingly, the width of the magnetization plateau also scales linearly in $\delta$.

The frustrating diagonal exchange can change qualitatively the magnetic phase diagram of our model only in the rather special case of a ladder with ferromagnetic legs and ferromagnetic diagonal exchange. In this case the anisotropy parameter of the effective spin-chain model is given by
\be
\Delta  = - \frac{1}{2}\frac{J_{\parallel} - J^{\prime}_{\perp}}{J_{\parallel} + J^{\prime}_{\perp}}\, .
\label{eff_2} 
\ee
This relation implies that for  
\be
J^{\prime}_{\perp} < J^{\prime c}_{\perp} =- \frac{\sqrt{2}-1}{\sqrt{2}+1}|J_{\parallel}| 
\ee
the spin stiffness parameter $K$ exceeds 2. In this parameter range the sine-Gordon model is  gapless and the magnetization plateau disappears. 

\section{Conclusion}

We have studied the ground state phase diagram of a spin $S=1/2$ two-leg 
ladder with alternating rung-exchange $J^{n}_{\perp} = J_{\perp}\left[1 + (-1)^{n} \delta \right]$ in a magnetic field. We have shown that in a wide parameter range 
the magnetization curve exhibits a plateau at one-half of its saturation value. The width of the plateau, is proportional to the excitation gap in the system at $M=0.5M_{sat}$ and  scales as $\delta^{\nu}$. The critical exponent has a value $\nu =4/5$ in the case of a ladder with isotropic antiferromagnetic legs and $\nu =2 $ in the case of a ladder with isotropic ferromagnetic legs. We have also shown that in the ladder with frustrating diagonal interleg exchange the plateau effect is stronger and for realistic values of diagonal exchange the critical exponent reaches the value $\nu=2/3$.

We have also shown that in the case of ladder with ferromagnetic legs and with strong 
diagonal ferromagnetic intraleg interaction, the magnetization plateau $M=0.5M_{sat}$ is absent. 

To conclude, we briefly comment the possible spontaneous appearance of an alternating rung exchange as a spin-Peierls instability. Assuming that in the harmonic approximation the lattice deformation energy per rung is given by $E_{def} \sim \delta^{2}$ and estimating the magnetic condensation energy as $E_{mag}(\delta)-E_{mag}(0) \sim  -\delta^{2\nu}$, we conclude that such an instability is possible for an antiferromagnetic ladder for a magnetization $M=0.5M_{sat}$, i.e. for $H = H _{c} \simeq J_{\perp} + 0.5J_{\parallel}$.

\section{Acknowledgments} 
 
It is our pleasure to thank D. Baeriswyl, D. Cabra, G. Jackeli, F. Mila, H.-J. Mikeska, S. Mahdavifar and T. Vekua for fruitful discussions. GIJ acknowledges support through grant Nr.200020-105446 of the Swiss NSF and the generous hospitality of the Department of Physics of the University of Fribourg, where part of this work was done.
-------------------------------------------------------------------------

\end{document}